\documentclass[aps,pra,twocolumn,groupedaddress,showpacs,superscriptaddress]{revtex4-1}
\usepackage{graphicx,amsmath}
\usepackage{amssymb}

\begin{document}
\title{Determination of Chain Strength induced by Embedding in D-Wave Quantum Annealer}
\author{Hunpyo Lee}
\affiliation{Department of Liberal Studies, Kangwon National University, Samcheok, 25913, Republic of Korea}
\email{Email: hplee@kangwon.ac.kr}
\date{\today}

\begin{abstract}
The D-wave quantum annealer requires embedding with ferromagnetic (FM) chains connected by several qubits, 
because it cannot capture exact long-range coupling between qubits, and retains the specific architecture 
that depends on the hardware type. Therefore, determination of the chain strength $J_c$ required to sustain 
FM order of qubits in the chains is crucial for the accuracy of quantum annealing. In this study, we devise 
combinatorial optimization problems with ordered and disordered qubits for various embeddings to predict 
appropriate $J_c$ values. We analyze the energy interval $\Delta_s$ and $\Delta_c$ between ground and first 
excited states in the combinatorial optimization problems without and with chains respectively, using the 
exact approach. We also measure the probability $p$ that the exact ground energy per site $E_g$ is observed 
in many simulated annealing shots. We demonstrate that the determination of $J_c$ is increasingly sensitive 
with growing disorder of qubits in the combinatorial optimization problems. In addition, the values of 
appropriate $J_c$, where the values of $p$ are at a maximum, increase with decreasing $\Delta_s$. Finally, 
the appropriate value of $J_c$ is shown to be observed at approximately $\Delta_c/\Delta_s=0.25$ and $2.1 
E_g$ in the ordered and disordered qubits, respectively.
\end{abstract}

\maketitle

\section{Introduction\label{Introduction}}

The recent progress in quantum technology has brought about the dawn of quantum machines. Machines based on 
qubits rather than classical binary digits are being developed and built more frequently than 
ever~\cite{Preskill2018}. One such machine is the D-wave quantum annealer (DQA)~\cite{Johnson2011}. Unlike 
gate-type quantum machines using circuits, the DQA implements a quantum annealing (QA) process in the 
parameterized Hamiltonian of a transverse-field Ising model, composed of binary superconducting 
qubits~\cite{Johnson2011,Kadowaki1998}. The primary advantage of this architecture is that it is 
much easier to add qubits than that in the case of gate-type quantum computers while maintaining the 
accuracy of results~\cite{King2022}. Thus, the emerging DQA, which has seen rapid increments in qubit 
capacity, is catching up to the computational speed of the classical digit machine in an annealing process. 
Consequently, it has been extensively employed in combinatorial optimization problems requiring an annealing 
process as well as in the research of the Ising model, which shows unconventional phases at zero 
temperature~\cite{Amin2018, Isakov2016, Mazzola2017, Inoue2021, Kairys2020, King2021, Irie2021, Park2022, 
Ronnow2014, Albash2018, Ronnow2014(1)}.

However, the DQA cannot technically describe exact couplings between long distance qubits. It also retains 
specific architectures such as Pegasus and Kimera graphs dependent on DQA hardware types. These situations 
demand a physical embedding of the problem into the DQA, such that the architecture of the original problem 
topologically matches with one on the DQA~\cite{Lanthaler2021, Konz2021}. Additionally, the chain with 
ferromagnetic (FM) coupling $J_c$ between several qubits necessitates that one variable in the architecture 
of the original problem be introduced in the embedding. Therefore, $J_c$ energetically competes with the 
coupling given in the original combinatorial optimization problem. The weak or strong $J_c$ induce the 
brokenness or the clustering of chains, respectively, which lower the accuracy of results measured by the 
DQA. This can be improved by appropriate selection of $J_c$.

\begin{figure}
\includegraphics[width=1.0\columnwidth]{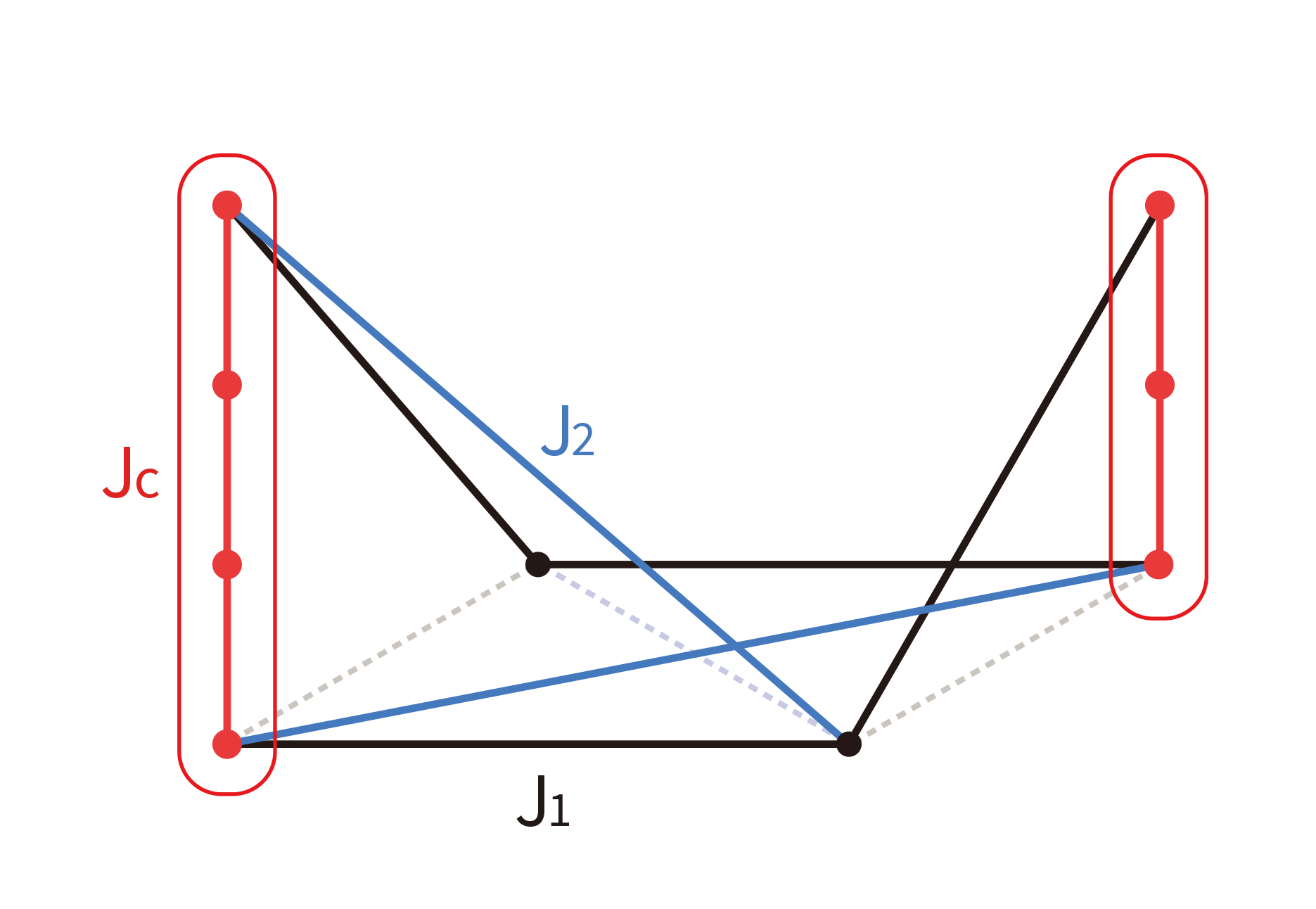}
\caption{\label{Fig1} (Color online) Schematic unit cell of $2 \times 2$ qubits on two-dimensional (2D) 
square structure with the nearest-neighbor interaction of $J_1$ and diagonal-neighbor 
interaction of $J_2$, as combination optimization problem. The chains with coupling of $J_c$ are made 
on three-dimensional $z$-direction of 2D architecture.}
\end{figure}

In this study, we devise artificial ordered and disordered systems to estimate the appropriate $J_c$ between 
qubits in the chains appearing in various embeddings. The qubits for the combinatorial optimization problem  
are put on two-dimensional (2D) $L \times L$ square architecture. The chains are made on the 
three-dimensional (3D) $z$-direction of the initially 2D architecture. The qubits in the original problem of 
2D architecture are connected on edge qubits of the chains in the $z$-direction.
The devised systems cover all embeddings of 2D by adjustment of the composition of qubits in the chains. 
We consider the 2D $L \times L$ frustrated and disordered Ising model as the combinatorial optimization 
problem, where the intervals $\Delta_s$ between ground and first excited energies are systematically tuned 
by strength of frustration. Fig.~\ref{Fig1} shows the schematic unit cell of $2 \times 2$ qubits on two-
dimensional (2D) square structure with the nearest-neighbor interaction of $J_1$ and diagonal-neighbor 
interaction of $J_2$. The chains are marked as circles. We control various parameters such as the distance, 
position, and number of the chains to provide information of approximate $J_c$ in many cases. We analyze the 
energetic model without the constraint of the penalty function, through the exact and simulated annealing 
(SA) approaches~\cite{Kirkpatrick1983,Santoro2002}.

We calculate the full energy spectrum in 2D $4 \times 4$ ordered qubits with nine qubits in the chains using 
the exact method. We also analyze the probability $p$ happened exact ground energy per site $E_g$ in many SA 
shots in 2D ordered and disordered $L \times L$ qubits with several chain strengths of various embeddings. 
We find that the energy gap $\Delta_c$ between ground and first excited states in the combinatorial 
optimization problem with chains of DQA are systematically controlled by $J_c$ in the ordered qubits. We 
confirm that $J_c$ is less sensitive in the ordered combinatorial optimization problem with large 
$\Delta_s$, while it is highly sensitive in the disordered one with $\Delta_s \approx 0$. We confirm that the most appropriate $J_c$, with the maximum values of $p$ for stable QA, increases with decreasing 
$\Delta_s$. Finally, we find that it occurs at $J_c=\Delta_c/\Delta_s=0.25$ and $2 E_g$ in the ordered phase 
and disordered phase, respectively, with $\Delta_s \approx 0.0$.

\begin{figure}
\includegraphics[width=1.05\columnwidth]{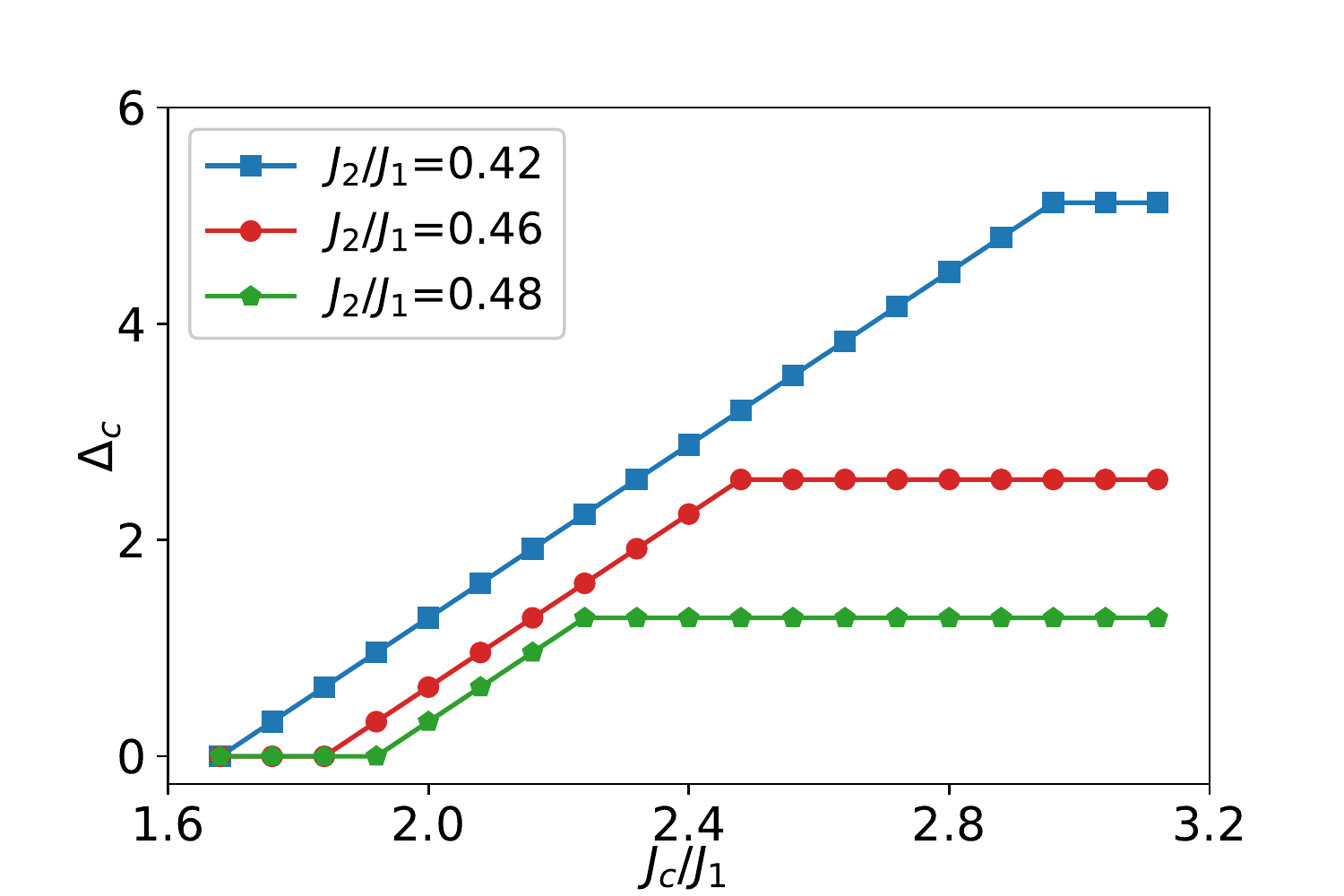}
\caption{\label{Fig2} (Color online) $\Delta_c$ between ground and first excited energies in pure 
$H_{\text{cop}}$ without the chains in Eq.~(\ref{Eq1}) for $J_2/J_1=0.42$, $0.46$ and $0.48$. $\Delta_c$ 
are computed by exact method with nine qubits in the chains on the 2D $4 \times 4$ 
qubits.}
\end{figure}

The paper is organized as follows: Section~\ref{model} gives a detailed description of the combinatorial 
optimization problem with ordered and disordered qubits for various embeddings. In Section~\ref{Result}, we 
predict $J_c$ for various parameters of the chains via exact and SA tools, and discuss results. Finally, we 
present the conclusions in Section~\ref{Conclusion}.

\section{Artificial combinatorial optimization model for Embedding\label{model}}

The Hamiltonian of the DQA is given as
\begin{equation}\label{Eq1}
H = H_{\text{cop}} + H_{\text{chain}},
\end{equation}
where $H_{\text{cop}}$ and $H_{\text{chain}}$ are the parts of the combinatorial optimization problem and 
the chain, respectively. Here, $H_{\text{chain}}$ is expressed as
\begin{equation}\label{Eq2}
H_{\text{chain}} = -J_c \sum_{i} \sum_{<k,k'>}^{n_i} \sigma_{i,k}^{z}\sigma_{i,k'}^{z},
\end{equation}
where $J_c$ means the chain coupling of FM order between $k$ and $k'$ qubits and $n_{i}$ is the number of 
qubits at the chain of $i$-site. The total number of qubits $N_c$ in all chains is given as $N_c=\sum_{i} 
n_{i}$. Unlike the realistic DQA with a specific architecture of qubits such as the Pegasus graph, we put 
the qubits of the combinatorial optimization problem and of the chains in 
$H_{\text{cop}}$ on a 2D $L \times L$ square lattice and in $H_{\text{chain}}$ in the
3D $z$-direction, respectively. These architectures cover all embeddings of 2D by control of the composition 
and number of qubits in the chains. To account for various embeddings, $i$ and $n_i$ are randomly selected 
in the $L \times L$ and in the size of $L$, respectively.

\begin{figure}
\includegraphics[width=1.05\columnwidth]{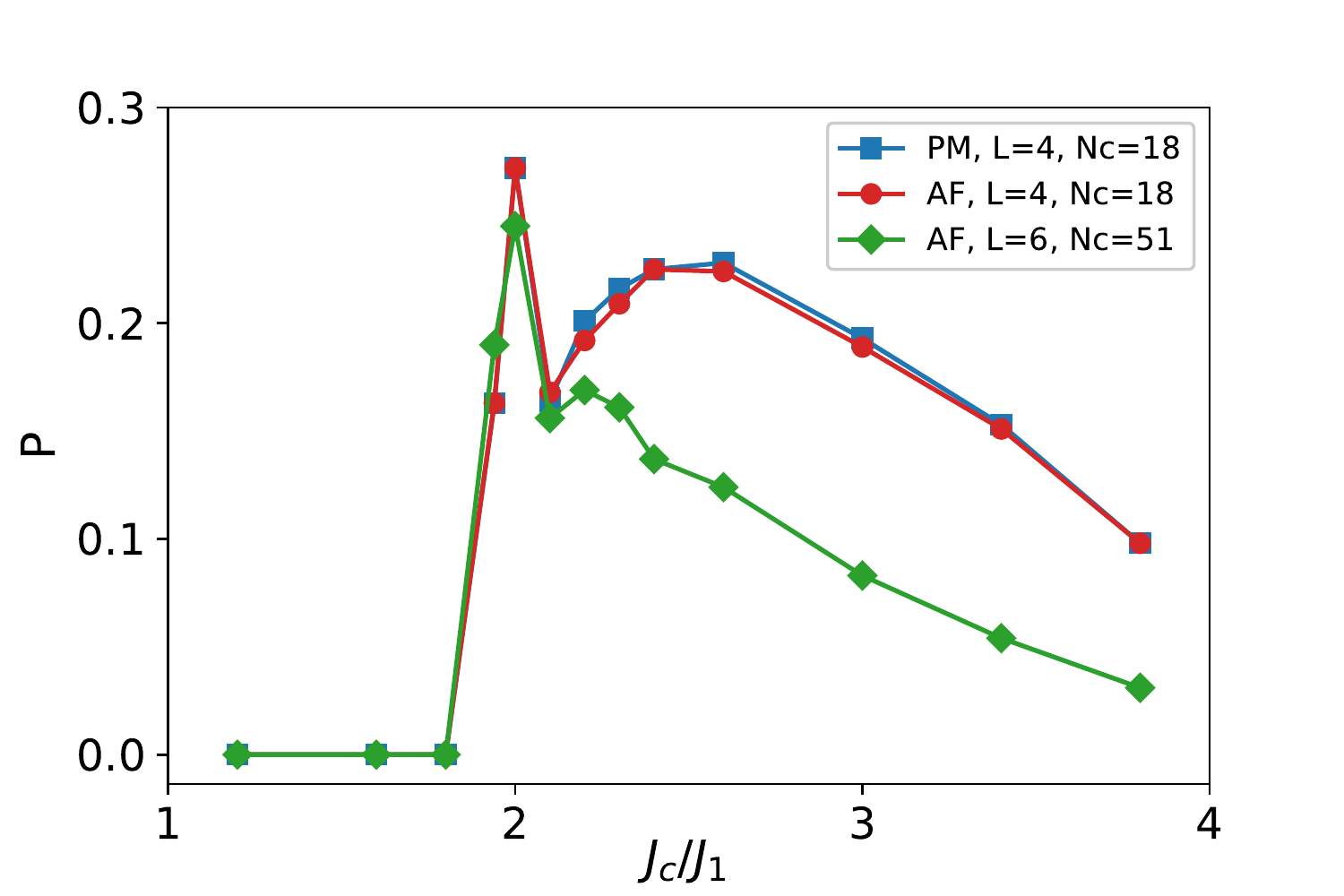}
\caption {\label{Fig3} (Color online) Probability $p$ that the exact ground energy per site is observed for 
$L=4$ and $6$ with paramagnetic (PM) and antiferromagnetic orders of qubits. This is done in many 
simulated annealing shots, where the positions and distances of the qubits in the chains are randomly 
selected in $i$-site on the ordered qubits and on $z$-direction of $i$-site, respectively. The critical 
chain strength $J_c^*/J_1$ and $J_c^{**}/J_1$ occurred the brokenness and clustering of the chains are 
$1.84$ and $2.48$, respectively.}
\end{figure}

To estimate appropriate $J_c$, as a simple example we consider the combinatorial optimization problem with 
the frustrated and disordered qubits tuned by diagonal couplings. The Hamiltonian $H_{\text{cop}}$ of the 
combination optimization problem is defined as  
\begin{equation}\label{Eq3}
H_{\text{cop}} = -J_1 \sum_{<i,j>} \sigma_i^z \sigma_j^z - J_2 \sum_{<<i,j'>>} \sigma_i^z \sigma_{j'}^z,
\end{equation}
where nearest- and diagonal-neighbors are denoted by $<i,j>$ and $<<i,j'>>$, respectively. The qubits of 
Eq.~(\ref{Eq3}) without chains display the antiferromagnetic (or PM) and stripe orders for $J_2/J_1<0.5$ and 
$J_2/J_1>0.5$ at zero temperature, respectively~\cite{Jin2012,Jin2013}. The value of $J_2$ systematically 
tunes $\Delta_s$. Note that another gap $\Delta_c$ appears in the full DQA Hamiltonian in Eq.~(\ref{Eq1}), 
where $J_c$ controls $\Delta_c$. The total number of qubits used in the DQA computation is $L^2+N_c$. The 
total ground energy of Eq.~(\ref{Eq1}) is given as $L^2 E_g + N_c J_c$.

\section{Result\label{Result}}

We first search for the critical chain strength $J_c^*/J_1$ occurred the chain brokenness in the 2D $4 
\times 4$ optimization problem with increasing the distance and number of the chains through exact approach. 
Surprisingly, we confirm that $J_c^*/J_1$ is not dependent on those of the chains in the regions of $-J_2/
J_1<0.5$ (or $J_2/J_1<0.5$) with antiferromagnetic (or PM) order of qubits. Fig.~\ref{Fig2} shows $\Delta_c$ 
as a function of $J_c/J_1$ for $J_2/J_1=0.42$, $0.46$ and $0.48$. Two kinks are evident in Fig.~\ref{Fig2}. 
$J_c^*/J_1$ with the first kink at $\Delta_c = 0$ are $1.68$, $1.84$ and $1.92$ for $4.2$, $4.6$ and $4.8$, 
respectively. $J_c^*/J_1$ increases with increasing $J_2/J_1$. The second kink when $J_c^{**}/J_1$ 
are $2.96$, $2.48$ and $2.24$, appears $0.42$, $0.46$ and $0.48$, respectively. $\Delta_c$ of these 
$J_c^{**}/J_1$ are exactly equal to $\Delta_s$ calculated by Eq.~(\ref{Eq3}) without the chains. We propose 
that the brokenness and clustering of the chains would appear below $J_c^*/J_1$ and above $J_c^{**}/J_1$, 
respectively. The regime between $J_c^*/J_1$ and $J_c^{**}/J_1$ expected the stable QA computations is 
shrinking with increasing $J_2/J_1$. This relationship breaks down at the combinatorial optimization problem 
of fully frustrated qubits with $J_2/J_1=0.5$.

\begin{figure}
\includegraphics[width=1.05\columnwidth]{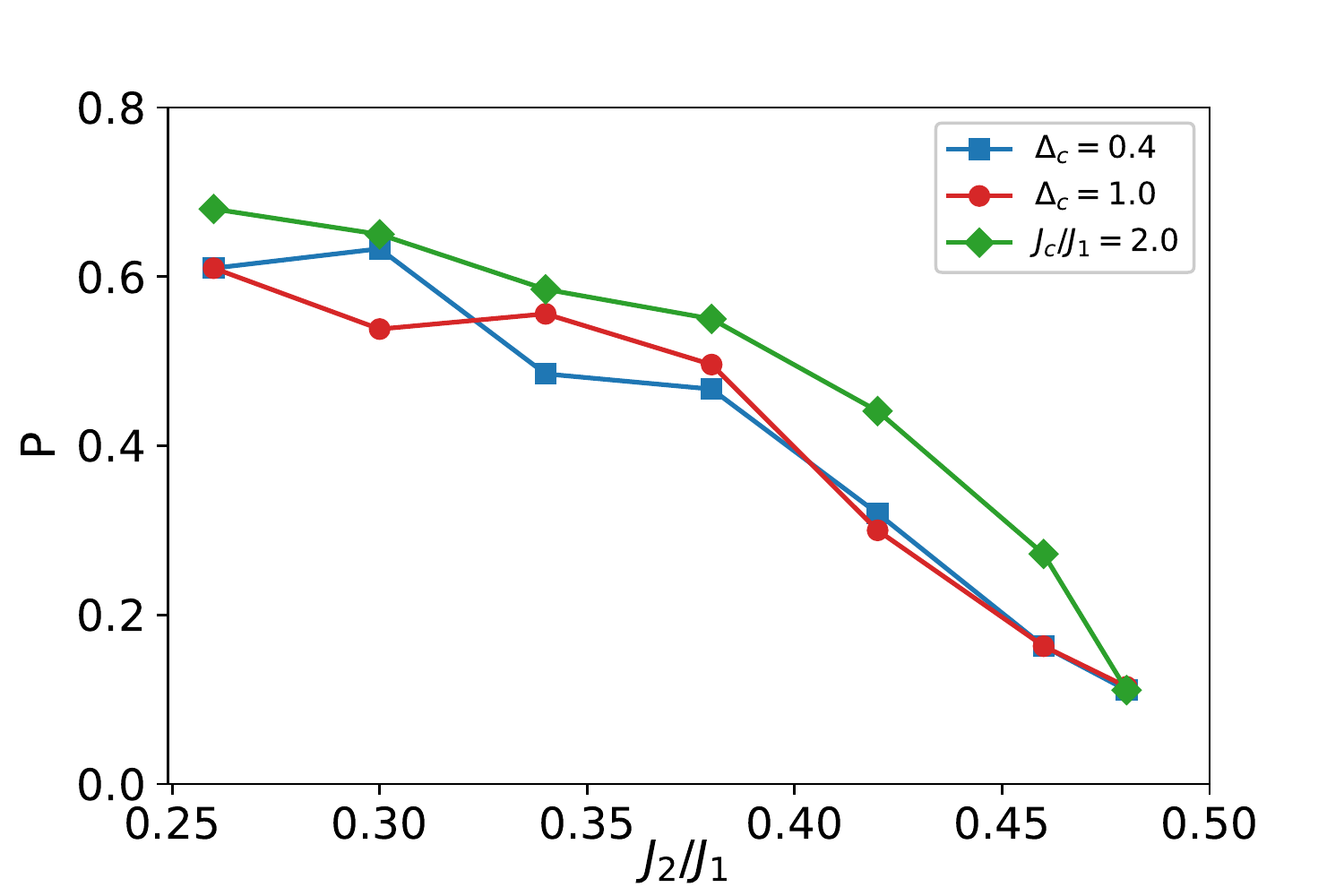}
\caption {\label{Fig4} (Color online) $p$ as a function of $J_2/J_1$ for $\Delta_c=0.4$, 
$1.0$ and $J_c/J_1=2.0$ for $L=4$. Here, in terms of energy gaps, $J_c/J_1=2.0$ is exactly equal to 
$\Delta_c/\Delta_s=0.25$ in the ordered antiferromagnetic (or PM) regions.}
\end{figure}

Next, we analyze the probability $p$ that $E_g$ is observed. This is done in many SA shots to confirm the 
validity of the exact results calculated above. Fig.~\ref{Fig3} displays $p$ as a function of $J_c/J_1$ for 
$L=4$ and $6$ with PM and antiferromagnetic orders of qubits. PM and antiferromagnetic orders are observed 
at $J_2/J_1=0.46$ and $-0.46$, respectively. The positions and distances of the qubits in the chains are 
randomly selected in $i$-site on the 2D $L \times L$ qubits and on $z$-direction of $i$-site, respectively. 
$N_c$ used in Fig.~\ref{Fig3} is $18$ and $51$ for $L=4$ and $6$, respectively. As expected, in all cases, 
$p$ is zero at the brokenness state of the chains below the value of $J_c^*/J_1=1.84$ predicted by the exact 
results. Note that we do not insert $p$ of small values with the chain brokenness, which observe $E_g$ in 
the first or second excited states of the SA computations, in Fig.~\ref{Fig3}. $p$ gradually decreases in 
the clustering state of the chains above $J_c^{**}/J_1=2.48$. The values of $p$ at PM order with $L=4$ are 
equal to those at antiferromagnetic order within numerical deviations. Overall, $p$ decreases with 
increasing $L$ and $N_c$. The highest values of $p$ are observed at $J_c/J_1=2.0$ in all cases.

In the following section, we investigate why the highest peaks of $p$ are observed at $J_c/J_1=2.0$ in the 
stable QA region between the brokenness and clustering regions. We plot $p$ as a function of $J_2/J_1$ for 
$\Delta_c=0.4$, $1.0$ and $J_c/J_1=2.0$ in Fig.~\ref{Fig4}. Here, in terms of energy gaps, $J_c/J_1=2.0$ is 
equal to $\Delta_c/\Delta_s=0.25$ in the antiferromagnetic (or PM) regions. The values of $p$ in all cases 
are decreasing with increasing $J_2/J_1$. They at $J_c/J_1=2.0$ are always higher than those at 
$\Delta_c=0.4$ and $1.0$, even though they converge at $J_2/J_1=0.48$ with a tiny $\Delta_s$. We surmise 
that $\Delta_c/\Delta_s=0.25$ is the optimized point without any bias between energy of the combinatorial 
optimization problem and the chain in Eq.~(\ref{Eq2}) and Eq.~(\ref{Eq3}), respectively.

\begin{figure}
\includegraphics[width=1.05\columnwidth]{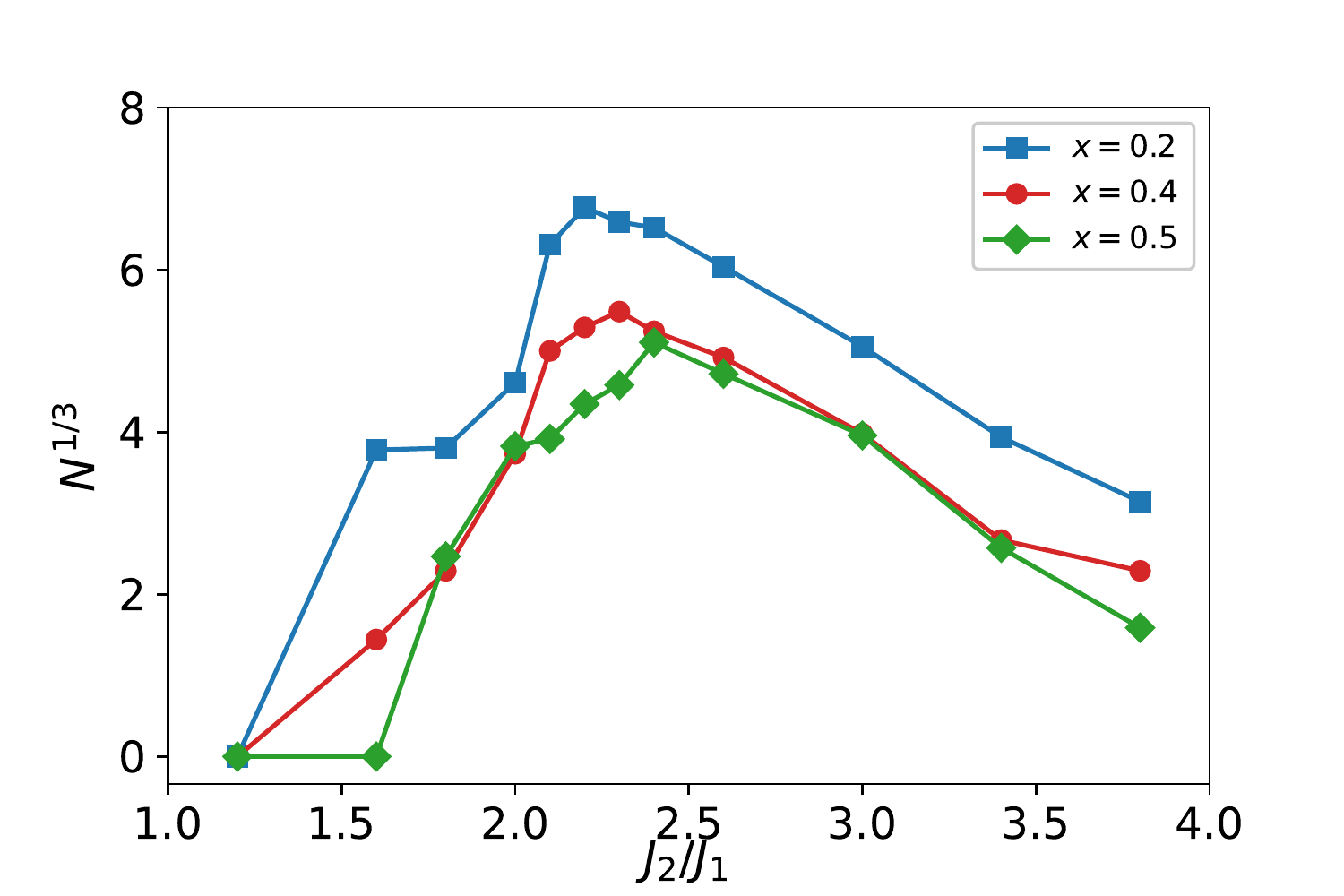}
\caption {\label{Fig5} (Color online) Number of cases $N$, where $E_g$ is observed in 2000 times SA 
shots, at $L=8$ for $x=0.2$, $0.4$ and $0.5$. Here, $x$ is a ratio to select $J_2=0.25$ and $1-J_2$ in the 
diagonal bond of Eq.~(\ref{Eq3}) to compose the disordered qubits.}
\end{figure}

Finally, we would like to search for appropriate $J_c$ in the realistic DQA, because most combinatorial 
optimization problems would be described using disordered qubits, while the above results use ordered 
qubits. For disordered systems, we consider the combinatorial optimization problem with competition between 
antiferromagnetic and stripe orders of the qubits. For those, we introduce an $x$ ratio to select $J_2=0.25$ 
and $1-J_2$ in the diagonal bond of Eq.~(\ref{Eq3}). For instance, $J_2$ and $1-J_2$ are selected according 
to the same ratio of $x=0.5$ in the maximally disordered states of the qubits. We count the number of cases 
$N$ where $E_g$ is observed in 2000 times SA shots. Fig~\ref{Fig5} shows $N^{1/3}$ as a function of 
$J_c/J_1$ at $L=8$ for $x=0.2$, $0.4$ and $0.5$. The ordered and disordered qubits first appear at $x=0.2$ 
and $0.4$, respectively. The maximally disordered qubits are seen at $0.5$. The averaged energy $<E_g>$ of 
ensembles is $-1.305$, $-1.149$ and $-1.117$ for $x=0.2$, $0.4$ and $0.5$, respectively. As 
expected, the number of cases where $E_g$ is observed as a function of $J_c/J_1$ is much larger in the 
ordered qubits with $x=0.2$ than in the disordered ones with $x=0.4$ and $0.5$. $J_c$ appeared the highest 
number of cases are $2.1$, $2.3$ and $2.4$ for $0.2$, $0.4$ and $0.5$, respectively. This means that the 
determination of appropriate $J_c$ is increasingly sensitive with growing disorder and that the appropriate 
value of $J_c$ increases with increasing disorder. We guess that the appropriate $J_c$ is observed at 
approximately $2.1 E_g$ in the disordered qubits.

\section{Conclusion\label{Conclusion}}

The DQA, with its recent rapid increase in qubit capacity, displays much potential and has been extensively 
applied in the solution of various combinatorial optimization problems. However, as a limitation, the 
physical embedding with FM chains of several qubits is required to consider exact long-range coupling 
between qubits ignored in the DQA. Brokenness and clustering of qubits in the chains lowers the accuracy of 
results measured by the DQA. To minimize this, appropriate determination of $J_c$ to keep the FM ordered 
qubits in the chains is crucial.

We designed the ordered and disordered qubits on the 2D $L \times L$ square structure as the combinatorial 
optimization problem. The chains with FM ordered qubits were composed on the 3D $z$-direction of the 2D 
architecture for examination of various embeddings. We used the exact method to compute the $J_c^*$ and 
$J_c^{**}$, at which chain brokenness and clustering happened, respectively. We analyzed the probability $p$ 
that $E_g$ occurred in the combinatorial optimization problems with ordered and 
disordered qubits to estimate appropriate $J_c$ in various embeddings through the SA approach. We found that 
$J_c$ is less sensitive in the ordered qubits with large $\Delta_s$, while it is highly sensitive in the 
disordered ones with $\Delta_s \approx 0$. In addition, we found that the most appropriate $J_c$, with the 
maximum values of $p$, increases with decreasing $\Delta_c$. Finally, we confirm that the appropriate $J_c$ 
is found at approximately $\Delta_c/\Delta_s=0.25$ and $2.1 E_g$ in the ordered and disordered qubits with $
\Delta_s \approx 0.0$, respectively.

We would like to note that before this work, we measured the qubit configurations of Eq.~(\ref{Eq3}) using 
the DQA with 5000+ qubits composed on the Pegasus graph~\cite{Park2022}. In our experience the QA on 
hardware shows stronger chain brokenness than SA on classical machine. Therefore, $E_g$ occasionally 
occurred within the chain brokenness phase, which is smaller than $J_c^*$. Nevertheless, the overall 
tendency of $p$ measured by DQA is qualitatively consistent with that of $p$ computed by the SA approach.

\section{Acknowledgements}
We would like to thank Hayun Park and Myeonghun Park for useful discussions. This work was supported by 
Ministry of Science through NRF-2021R1111A2057259. We acknowledge the hospitality at APCTP where part of 
this work was done.

\end{document}